\documentclass[aps,prb,groupedaddress]{revtex4}
\usepackage{graphicx}
\usepackage{epstopdf}
\usepackage[margin=10pt,font=small,labelfont=bf]{caption}

\begin{document}

\title{Blogging in the physics classroom: A research-based approach to shaping students' attitudes towards physics}

\author{Gintaras D$\bar \textrm{u}$da}
\email{gkduda@creighton.edu}
\author{Katherine Garrett} \email{KatherineGarrett@creighton.edu} \affiliation{Department of
Physics, Creighton University, 2500 California Plaza, Omaha, NE
68178, USA}

\begin{abstract}
  \vbox{Even though there has been a tremendous amount of research done in how to help students learn physics,
  students are still coming away missing a crucial piece of the puzzle: why bother with physics?  Students
  learn fundamental laws and how to calculate, but come out of a general physics course without a deep understanding
  of how physics has transformed the world around them. In other words, they get the ``how" but not the ``why".  Studies
  have shown that students leave introductory physics courses almost universally with decreased expectations and with a more
  negative attitude.  This paper will detail an experiment to address this problem:  a course weblog or ``blog" which discusses real-world
  applications of physics and engages students in discussion and thinking outside of
  class. Specifically, students' attitudes towards the value of physics and its applicability to the real-world were probed using a 26-question Likert scale
  survey over the course of four semesters in an introductory physics  course at a comprehensive Jesuit university. We found that
  students who did not participate in the blog study generally exhibited a deterioration in attitude towards physics as seen previously.
  However, students who read, commented, and were involved with the blog maintained their
  initially positive attitudes towards physics.  Student response to the blog was overwhelmingly positive, with students
  claiming that the blog made the things we studied in the classroom come alive for them and seem much more relevant.}
\end{abstract}

\maketitle

\section{Introduction}

An introductory physics course has many goals, some of which are
explicitly stated in the syllabus and some of which are a bit more
elusive.  A great deal of progress has been made in recent years
on these explicit goals -- improving student learning and
conceptual understanding in introductory physics and astronomy
courses \cite{McDermott,Redish1,Redish2,Redish3}. Physics
education research has led to major improvements in student
comprehension and learning by focusing on active rather than
passive learning strategies in the classroom using a wide variety
of diverse approaches such as Peer Instruction (PI)\cite{Mazur},
Just in Time Teaching (JITT)\cite{JITT}, University of Washington
style tutorials\cite{UofW}, Workshop Physics\cite{workshop}, and
many others.  However, those more elusive goals, aptly described
as belonging to the ``hidden curriculum"\cite{MPEX} remain
difficult to address.

One goal which we believe is important to most, if not all,
instructors and which clearly falls into the ``hidden curriculum"
category is to leave students with a positive attitude towards
physics.  This goes beyond simply ``liking" physics; it
encompasses an appreciation of how physicists think and operate,
the value of physics as it applies to  other fields such as
engineering, biology, medicine, etc., and the applicability of
physics to everyday life.  To use JITT terminology, instructors
would like students to see what physics is ``good-for."  Although
we believe this is an important goal, physicists (and astronomers)
have had little success addressing this dimension in the
classroom.

In response to this concern, Redish, Saul, and Steinberg designed
the Maryland Physics Expectations (MPEX) Survey, a 34-item Likert
scale survey that examines ``student attitudes, beliefs, and
assumptions about physics" \cite{MPEX}. The intent was to study
how introductory physics students' attitudes and expectations
changed due to taking a physics course. As they write:

\begin{quote}
``In all cases, the result of instruction on the overall survey
was an \textit{increase} in unfavorable responses and a decrease
in favorable responses ... \textit{Thus instruction produced an
average deterioration rather than an improvement of student
expectations."}
\end{quote}

\noindent In other words, students' attitudes towards physics were
more negative after taking an introductory physics course.  This
should certainly be a source of dismay to physics teachers.

Zeilik, Schau, and Mattern observed the same phenomenon in
introductory astronomy courses \cite{Zeilik}. Using data from over
400 students at the University of New Mexico in introductory
astronomy courses they concluded that there was ``little change
over each semester in students' mildly positive incoming attitudes
about astronomy and science." This change in attitude occurred
despite the use of innovative assessment techniques such as
concept maps, small group work, and identification of student
misconceptions \cite{Zeilik2}.  In other words, re-organizing a
class to emphasize active-learning and employing novel assessment
techniques did not seem to affect student attitudes towards
astronomy.

More recently Adams et al. have developed a new instrument for
measuring student attitudes and beliefs in introductory physics
courses over a wide range of categories such as personal interest,
real world connections and sense making, called the Colorado
Learning Attitudes about Science Survey (CLASS)\cite{CLASS}.  They
found that, ``most teaching practices cause substantial drops in
student scores".  This phenomena is not unique to physics. Other
disciplines, such as chemistry and statistics, have similarly
begun to seriously examine the role and effect of students'
attitudes in their respective fields
\cite{stats1,stats2,chem1,chem2}.  In fact, when the CLASS
instrument was modified for use in introductory chemistry courses
at a large state research university the ``results indicated that
shifts after instruction were similar to, if not worse than, in
physics in moving in the unfavorable direction" \cite{CLASS}.

Why, however, should we worry about student attitudes?  Most
instructors aim to create an outgoing positive attitude towards
physics, and educational research has repeatedly shown over the
past few decades that learning is intrinsically linked with
student attitude and expectations \cite{Schoenfeld,Koballa}. To
cite just one specific example, Coletta and Philips (at Loyola
Marymount) claim to have found a strong correlation between FCI
gains \cite{FCI} and MPEX scores for their students at Loyola
Marymount University\cite{Coletta}. In other words, if we care
about learning we need to pay attention to students' attitudes.
The goal of this paper is to explore whether or not there is a way
to positively impact student attitudes in an introductory physics
course.  To this end, we have examined the effectiveness of a
course weblog or ``blog" in shaping and guiding students'
attitudes (see Blood's essay \cite{blogs1} for a primer on blogs
in general).

But why use a blog?  Obviously, blogs are an example of new
technology which in and of itself may be appealing to increasingly
technologically savvy students. However, technology should never
be employed simply for the sake of showcasing technology.  We use
a blog due to the wealth of recent pedagogical research that has
shown the value and validity of using blogs in both science and
non-science courses
\cite{blogs2,blogs3,blogs4,blogs5,blogs6,blogs7}. For example,
Ferdig and Trammel present some compelling arguments for blogging
in the classroom, enumerating four primary benefits to student's
involvement in blogs\cite{blogs6}: (1) when writing posts or
comments students must scour through vast amounts of information
on the web or in other references; this not only exposes them to a
broad range of topics outside the classroom, but it also forces
them to evaluate the validity and the value of various sources,
(2) blogging tends to increase student excitement for learning and
ownership of the process,  (3) blogs open up discussions to
students who may not otherwise participate in class, and (4)
blogging encourages discussion outside of class with a wide
variety of viewpoints.

Ferdig and Trammel also point out that, ``current educational
research and theory have demonstrated the importance of social
interaction in teaching and learning"\cite{blogs6}; blogs provide
a way for students to interact with each other and their
instructor outside of class.  Halavais also stresses the ability
of blogs to move student learning outside the classroom
\cite{blogs7}.   And finally, Brownstein and Klein report that
since the introduction of blogs in their classroom ``the focus was
moved from `what' to `why'", and that their students ``see
knowledge as interconnected as opposed to a set of discrete facts"
\cite{blogs8}.  Blogs appear to be a quite promising tool for
addressing the ``hidden curriculum" that we consider essential but
that has been frustratingly elusive to pass on to our students.

Since this is such a large and open-ended question, as a first
step we limit our focus on the effect on student attitude; the
effect on student learning is left to be considered in the future.
As our operational definition of attitude we use the MPEX concept
of expectations, an all-inclusive set of beliefs, attitudes,
skills, views on science and physics in general, and the value of
physics \cite{MPEX}.  This correlates with what the CLASS
instrument refers to as ``beliefs" \cite{CLASS}.  We restrict our
study of attitude to a sub-section of expectations/beliefs
revolving around students' value of physics as it relates to their
careers and the real-world, which the MPEX survey refers to as the
``reality-link" category or the ``real-world connections" category
in the CLASS instrument.

Apart from the question of student attitudes in introductory
physics classes, we hope that this study provides an example of
how to apply the new technology of blogging to the physics
classroom.

\section{Course Implementation}

General physics at Creighton University (CU) is taught in three to
four sections of approximately 30 students meeting three times a
week for 50 minute lectures, once a week for a one hour recitation
section, as well as a two hour laboratory session.  Traditional
lectures are supplemented with active learning strategies such as
Peer Instruction (PI) \cite{Mazur} and elements of Just in Time
Teaching (JITT) \cite{JITT} with recitation frequently making use
of materials developed through physics education research such as
ranking tasks and University of Washington style tutorials
\cite{UofW}. The corresponding laboratory section employs a
mixture of Real Time physics and Workshop physics elements.
Approximately 60-70\% of the general physics students at CU are
interested in the health professions (medical school, dental
school, physical therapy, etc.).  Due to local medical school
entrance requirements (CU as well as the University of Nebraska),
the general physics sequence is calculus-based, and pre-health
students take the same sequence as physics majors and other
physical science students.  This student composition is a unique
feature of physics at CU, and makes addressing the cross-utility
of physics in other fields essential; one can easily imagine that
engineers and scientists are far more likely to see the utility of
physics than students focusing on biology and chemistry in
preparation for a career in medicine.

This study was conducted over the course of four semesters as
summarized below in Table~\ref{tab:1}.  In Semester I (Fall 2005)
two sections, A and B, participated in the study while the
remaining two sections, sections C and D, served as a control
group.  In Semester II (Spring 2006) the blog was extended to all
four sections in part due to overwhelmingly positive participating
student feedback and in part due to demand from students who did
not participate during Semester I.  In Semester III (Fall 2006)
all four sections again participated in the study, and in Semester
IV (Spring 2007) only one section was included (mainly due to
instructor turnover).

\begin{table}[h]
\begin{center}
\begin{tabular}{llll}
\hline \hline Semester & Date & Instructors & Participated \\
& (number & & in blog \\
& participant) & & study \\
\hline Semester I & Fall 2005 &  &  \\
& (n=31) & Instructor A (GD) & yes \\
& (n=27) & Instructor B & yes \\
& (n=32) & Instructor C & no (control) \\
& (n=24) & Instructor D & no (control) \\
Semester II & Spring 2006 & & \\
& (n=36) & Instructor A (GD) & yes \\
& (n=35) & Instructor B & yes \\
& (n=33) & Instructor C & yes \\
& (n=25) & Instructor D & yes \\
Semester III & Fall 2006 & & \\
& (n=33) & Instructor B & yes \\
& (n=34) & Instructor C & yes \\
& (n=30) & Instructor D & yes \\
& (n=28) & Instructor E & yes \\
Semester IV & Spring 2007 & & \\
& (n=33) & Instructor A (GD) & yes \\
\hline \hline
\end{tabular}
\end{center}
\caption{Summary of Semesters I-IV.} \label{tab:1}
\end{table}

The blog was integrated into the course as follows: since reading
the blog would be on top of the numerous assignments, reading
quizzes, and exams that general physics students at CU already had
to complete, we decided to assign the blog as extra credit.   The
course instructors (mainly GD) posted several times a week to the
course blog. Students received two points of extra credit per week
for (1) reading the posts to the course blog during the week and
(2) for posting comments to one or more posts (each two points
corresponded roughly to 2\% on an exam or 0.2\% of the overall
course grade; in other words, by participating in the blog every
week students could raise their overall grade by about 2.5\%). The
criteria for student comments were that they be a thoughtful and
articulate reflection on the blog post, about a paragraph in
length, that tied in outside information relevant to the topic in
question.  In other words, students had to move beyond a simple
``This is cool!" response and include some actual content, much of
which was the result of additional research on their part.

In terms of instructor workload, the average time spent on the
blog (writing posts, responding to comments, recording credit) was
approximately two to three hours per week.  However, posting to
the blog did not necessitate always re-inventing the wheel.  For
example, the website \textit{How Stuff Works} \cite{howstuffworks}
graciously gave permission to use their content on the blog, as
long as it was for educational purposes, properly cited, and
contained links which pointed back to their webpage. Quick posts
at times were little more than an introductory paragraph followed
by a link to a interesting article on, for example, the physics of
television or the physics behind photocopying.  In other words,
weekly or bi-weekly blog posts did not have to be marathon
projects, and often it was the short, link-type posts which
students enjoyed the most.  Posts were re-used (often with some
improvement based on student input) between semesters (student
comments were deleted so as to start with a blank slate); in this
manner a ``library" of blog posts was built up that could be
rapidly and easily used in the future.

Since two to three hours per week is not a small investment of
instructor time, a few additional comments on instructor workload
are in order.  Once a library of blog posts has been built up, as
previously mentioned, the instructor time commitment drops to the
time necessary to read student comments, record their
participation, and answer questions on the blog.  For a class of
about 120 students in total at Creighton (with about 75\%
participating in the blog) we estimate that this entails roughly
thirty minutes of work per week.  Teaching assistants can be used
to monitor the blog, answer student questions, and record
participation.  In this way, the blog is easily scalable to larger
classes (however, more blog posts per week would be required to
limit saturation of student comments or an individual blog could
be created for each section of the course).  A complete set of
blog posts has been posted at the EPAPS depository to cut down on
the large initial investment of time for instructors interested in
using the blog.  And finally, we have had success in having
students write their own blog posts for the class to read; this
can involve some formatting work, however, students can be given
access to the blog to work as a collaborative team.

Tables~\ref{tab:2} and \ref{tab:3} below give a partial list of
general physics topics with their corresponding blog posts (a
version of the course blog from Semester II without student
comments is still available for viewing\cite{cugenphys} and
Appendix II contains a blog post from Semester I with several
actual student comments). Blog posts also integrated a wide
variety of physics applets and videos available on the web. In
particular, YouTube\cite{YouTube} proved to be a valuable resource
which students were quite familiar with (and excited to see used
in the classroom). For example, one blog post included a link to a
YouTube video of a car being struck by lightning with the
passengers none the worse for the experience; the video definitely
caught students' attention and led to a natural discussion of
Faraday cages (as well as students' critiques of the explanations
provided on the YouTube comments for the video).

The content of the blog mainly focused on how the physics we were
currently studying applied to the ``real world" and other fields
besides physics.  The individual posts were written in the same
theme as a JITT ``Good For"\cite{Good For}, however there are some
important differences. A physics ``Good For" entails private
communication between an instructor and an individual student
whereas student responses to blog posts are public.  Students
respond in an open forum, react to each other's comments, and pose
and answer questions. This had the effect of moving discussion out
of the classroom.  Whereas a physics ``Good For" poses a specific
list of questions to answer for credit, the blog post responses
tended to be more open and flexible; students could use the post
and bring in related topics that were of particular interest to
them. Students interested in biology would often comment with
biological examples; for example when discussing an application of
static electricity a biology student discussed the importance of
electrostatics in insect pollination. The blog was updated more
frequently (2-3 posts per week), and tended to focus on topics
specifically tailored to interest our health-science heavy
classes.

For semesters I-III, Blogger \cite{blogger}, a free blog-hosting
service, was used for the course blog.  Haloscan \cite{haloscan},
a free comment package for blogs, was used due to additional
functionality over Blogger comments.  Sitemeter was utilized to
keep track of visits to the blog, particularly to make sure that
CU general physics students were the primary visitors to the blog.
By Spring of 2007 (Semester IV) CU had implemented its own
blogging service utilizing WordPress \cite{wordpress}, hence the
blog in Semester IV was locally hosted.  In terms of pedagogy and
utilization it made little difference whether using Blogger or
WordPress (other than the learning curve that the switch entailed
for the instructor).

\begin{table}[h]
\begin{center}
\begin{tabular}{ll}
\hline \hline Physics Subject Area & Blog Topic \\
\hline
1D Kinematics & Speed and Acceleration \\
Vectors & GPS System and Triangulation \\
2D Kinematics & Projectile Motion and Air Resistance \\
Uniform Circular Motion & Centrifuges and Carnival Rides \\
Newton's Laws & Physiological Effects of Acceleration \\
& The Magnus Force \\
& The Equivalence Principle \\
& Friction: How Geckos scale walls \\
& Non-inertial frames: Coriolis Effect \\
& Atomic Force Microscopy and nN \\
Energy & Einstein's Famous E = mc$^2$ \\
& The Physics of \textit{Armageddon} \\
& Energy Use and your Body \\
Fluid Mechanics & Archimedes and King Hieron's Crown \\
& The Venturi Effect \\
& Physics of Blood Flow \\
Conservation of Momentum & Collisions \\
Angular Momentum & Pulsars and Neutron Stars \\
& Torques and your Body \\
Gravitation & Satellites \\
Oscillations/Resonance & The London Millenium Bridge \\
& The Foucault Pendulum \\
 \hline \hline
\end{tabular}
\end{center}
\caption{General Physics I topics and corresponding blog posts.}
\label{tab:2}
\end{table}

\begin{table}[h]
\begin{center}
\begin{tabular}{ll}
\hline \hline Physics Subject Area & Blog Topic \\
\hline Electrostatics & Static Electricity \\
& Physics of Lightning \\
& Photocopiers \\
& Electric Eels \\
Magnetism & The MagnaDoodle \\
& The Earth's Magnetic Field \\
& Magnets and Mystic Healing \\
& Physics of Television \\
Capacitance & Shampoo Bottle as Capacitor \\
& iPod Touchwheel \\
Induction/Faraday's Law & Stereo Speakers \\
& Credit Card Readers \\
& The Forever Flashlight \\
& Induction Furnaces \\
Waves & How Bent Spaghetti Breaks \\
& Water Waves and Tsunamis \\
& Bad Science in \textit{The Core} \\
Light and Optics & Mirages and Optical Illusions \\
& Solar Sails \\
& Polarization and Sunglasses \\
& Gravitational Lensing \\
& Lighthouses and Overhead Projectors \\
\hline \hline
\end{tabular}
\end{center}
\caption{General Physics II topics and corresponding blog posts.}
\label{tab:3}
\end{table}

\section{Attitudinal Survey: The Instrument}

To measure students' attitudes towards physics, we utilized a
5-point Likert scale survey. The attitudinal survey used in this
work is based on on the Attitude II survey designed by Zeilik,
Schau, and Mattern \cite{Zeilik} which was utilized in
introductory astronomy courses at UMN, and which was obtained from
the FLAG website \cite{FLAG}. The attitudinal survey was modified
to specifically address an introductory physics course (rather
than astronomy) with a few questions added.  Two versions of the
survey were given during the course of the semester: the pre-test
which was administered on the first day of class, and the
post-test which students completed at the end of the semester.

The attitudinal survey examines several dimensions: 1) affect, the
positive and negative attitudes students may have towards physics,
2) cognitive competence, student's confidence in their
intellectual skills and knowledge as applied to physics, 3) value,
the worth and relevance of physics to student's lives, and 4)
difficulty, a measure of the overall difficulty of physics to
students.  Our main modifications involved trimming the survey
down from thirty-four to twenty-six questions, mainly by removing
questions for the cognitive competence and difficulty categories.
The survey is included as Appendix I, which also cross-correlates
questions from this instrument to the MPEX and the CLASS tools and
discusses the reliability and validity of the instrument. As
mentioned in the introduction, particularly important to this work
is the value category, which corresponds well with the ``reality
link" category of the the MPEX survey and the ``real world
connections" category of the CLASS instrument. We feel that the
value category questions are similar enough to the corresponding
categories from the MPEX or CLASS instruments to allow for a
direct comparison with the trend of decreasing
attitude\cite{MPEX,CLASS}.  We have left a few items on cognitive
competence and difficulty; these, however, have not been analyzed
and will be considered in future work.

We chose to utilize the Attitude II survey of Zeilik et al. rather
than another instrument such as the MPEX, VASS\cite{VASS}, or
CLASS for several reasons.  First of all, the Attitude II survey
contained far more questions which probed students' sense of the
value of physics, precisely the dimension we were most interested
in; for example, the MPEX survey contains only four questions in
the ``reality link" category.  Since these surveys are Likert
scales and class-sizes at CU are much smaller than at larger,
public institutions, we felt that we would achieve more
statistically significant results by probing students' value of
physics with multiple questions.  Also, the CLASS instrument had
not yet been published when we began using the Attitude II survey
in Semester I (August, 2005).  However, as mentioned above, we
feel that the value items on the Attitude II survey match nicely
with the corresponding categories on the MPEX and CLASS
inventories.

\section{Data Analysis and Results}

Likert scale data is generally evaluated using two different
methods.  The first treats the survey results as interval data
(the differences between each response being equal) which allows
the data to be simply summed and statistically analyzed.  The
second treats the survey results as ordinal data, meaning that the
differences between each response are not equal and one should
therefore examine the percentage of students who agreed vs.
disagreed as the statistical measure. See the CLASS paper for an
excellent discussion of the difference between interval and
ordinal data \cite{CLASS}.  We evaluate the attitudinal data in
two different ways: 1) we group the value or ``reality link"
questions and perform a dependent-samples t-test and calculate an
effect size (treating the Likert survey results as interval data),
and 2) we use the agree-disagree plots and the binomial analysis
presented to highlight changes both for clusters of survey
questions as well as individual survey items (treating the survey
results as ordinal data). We concentrate on the second method
(treating the survey data as ordinal data) to compare our work to
previous work on the subject; for example, both the MPEX and CLASS
surveys are analyzed in this method as ordinal data.

Because students took the pre and post-tests anonymously in
Semester I and III (fall of 2005 and 2006) we compare
classes/sections as a whole. For Semester II (spring 2006) much
more extensive data were compiled allowing for a more detailed
statistical analysis.  All students in the course completed both
the pre and post attitudinal instrument.  Due to relatively small
class sizes, we have aggregated the data (in some cases by
sections in some cases by students who read and posted to the blog
versus those who did not) rather than reporting by each section to
improve the statistical significance of the results.

Agree-Disagree plots, or more simply A-D plots or ``Redish" plots,
were first presented by Redish, Saul, and Steinberg in analyzing
data from the MPEX Survey (treating the Likert scale survey
responses as ordinal data) \cite{MPEX}. Strongly and weakly agree
responses are combined together as are strongly disagree and
weakly disagree to form overall agree and disagree values.  These
values are then plotted usually as favorable versus unfavorable
responses (for some survey items disagreement is a favorable
result).  A straight line is then plotted from a 100\% favorable
to a 100\% unfavorable response; all student responses will fall
within in area bounded by the favorable and unfavorable axes and
this line.  A-D plots are useful in that they visually portray all
three dimensions of the data: the proportion of favorable and
unfavorable responses is evident from the position on the A-D
plot, and the proportion of neutral responses can be gauged from
the distance from the diagonal line described above.

To analyze the statistical significance of A-D plots Redish, Saul,
and Steinberg fit the data to a binomial distribution.  If $p$ and
$q$ are the percentages of students who answered favorably and
unfavorably respectively (not considering neutral responses), then
in order to fit a binomial distribution they introduce the
following renormalization:

\begin{equation}
p' = \frac{p}{p+q} ~~~~~ \textrm{and}~~~~ q' = \frac{q}{p+q}
\label{eq:1}
\end{equation}

They considered a shift in A-D values to be significant if the
difference was at less than the 5\% probability level.  In other
words, the shift must be greater than 2$\sigma$ where

\begin{equation}
\sigma = \sqrt{p'q'/n}, \label{eq:2}
\end{equation}

\noindent and $n$ is the number of students. Using their analysis,
for example, for values of $p=60$\% and $q=20$\% with $n=50$ we
consider a shift of 12\% to be statistically significant.

\subsection{Semesters I and III: Fall 2005 and 2006}

In Semester I the blog was offered as extra credit in two
sections: Sections A and B, taught respectively by Instructors A
(GD) and B.  Sections C and D did not participate in the blog
study to allow for a control group.  There was almost 100\%
participation in Section A due to encouragement by the instructor;
Instructor A (GD) frequently mentioned the blog in class and
reminded students to check and comment on the blog. Participation
in section B was roughly 60\%; Instructor B did not mention the
blog besides the initial description of the extra credit
available.  In Semester I data was not keyed to individual
students; students took the survey anonymously, although care was
taken to ensure students took both the pre and post tests.

\begin{figure}[htp]
\includegraphics[width=0.69\textwidth]{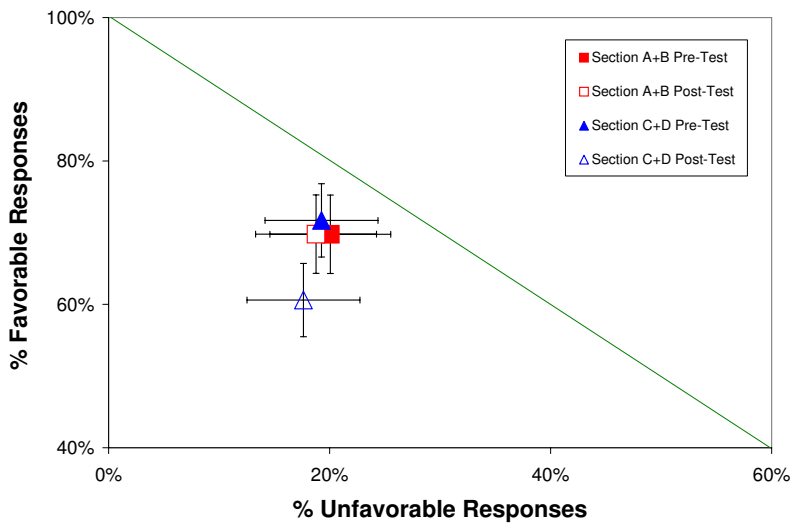}
\caption{A-D plot for Semester I (fall 2005) for sections which
read the blog (A+B combined) and sections which did not
participate in the blog study (sections C+D combined). Error bars
are 1$\sigma$, where $\sigma$ is calculated using Eq.(\ref{eq:2}).
In the binomial analysis a shift of $> 2\sigma$ is considered
statistically significant.} \label{fig:1}
\end{figure}

For Semester I, we take an average of the value or ``reality link"
questions for sections which participated in the blog study (A+B)
as well as for the two sections which did not (C+D) both for the
pre and post attitudinal survey.  We plot these averages on an
Agree-Disagree Redish plot as Fig.~\ref{fig:1}.  We found that for
the blog reading sections the agree-disagree values went from
69.7\% favorable and 20.0\% unfavorable on the pre-test to 69.7\%
favorable and 18.7\% unfavorable ont the post-test, a
statistically insignificant change.  However, for the two sections
which did not read the blog the agree-disagree values went from
71.7\% favorable and 19.3\% unfavorable on the pre-test to 60.6\%
favorable and 17.6\% unfavorable on the post-test.  For sections
C+D there was an 11.1\% decrease of favorable responses, which by
the binomial distribution criteria is a statistically significant
change.  Hence, students on average who read the blog had
maintained their initial positive attitude (with regards to the
value or ``reality link" aspect of physics) whereas students who
did not read the blog saw a general deterioration of their initial
positive attitude.

\begin{figure}[htp]
\includegraphics[width=0.69\textwidth]{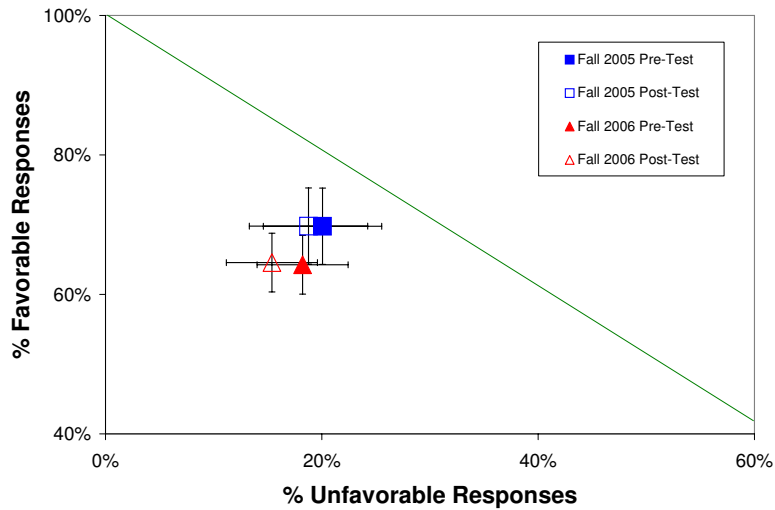}
\caption{A-D plot comparing students who participated in the blog
from Semester I (fall 2005) with students from Semester II (fall
2006).  Error bars are 1$\sigma$, where $\sigma$ is calculated
using Eq.(\ref{eq:2}). In the binomial analysis a shift of $>
2\sigma$ is considered statistically significant.} \label{fig:2}
\end{figure}

In Semester III, on the other hand, all four sections read and
commented on the blog.  In Semester III there was a slight change
of instructors; Instructor A (GD) was replaced by another
instructor due to a scheduled junior-faculty leave.  Instructor B
took over administrating the course blog.  Despite the change of
instructors, the averaged value or ``reality link" questions
showed the same trend as for fall 2005.  Students from Semester
III (who all had the opportunity to participate and read the blog)
maintained their positive attitudes as well as students from
Semester I who had read the blog.  The ``reality link" average for
Semester III went from 64.3\% favorable and 18.2\% unfavorable on
the pre-test to 64.6\% favorable and 15.4\% unfavorable on the
post-test.  The data for both semesters are plotted as
Fig.~\ref{fig:2}. This is one indication that the effect of the
blog is not instructor specific; the effect of ``instructor
immediacy", to use a communication studies phrase, will be
examined more thoroughly in a future work. We feel confident that
the student population from Fall 2005 to Fall 2006 (all CU
students with similar demographics such as major and class
standing) was similar enough to allow a direct comparison between
the two semesters.

\subsection{Semesters II and IV: Spring 2006 and 2007}

In Semester II the blog was offered as extra credit to all four
sections of general physics.  Instructor A (GD) was again
responsible for writing and maintaining the blog; blog topics were
frequently mentioned and discussed in section A.  To offset the
lack of a control group, responses on the attitudinal surveys were
recorded for each individual student, allowing their attitudinal
survey responses to be compared to blog participation, course
grade, results on standardized assessment exams like the DIRECT
and CSEM exams, etc \cite{DIRECT,CSEM}.

A significant question to consider is who actually read the blog.
If only the `A' students were reading the blog, one might expect
that any change in attitude was the result of over-sampling highly
motivated students.  However, upon examining the data, we found
only a small correlation between blog participation and final
course grade (with a correlation coefficient of 0.37); however,
since blog participation acted as a source of extra credit the
actual correlation is smaller (with a correlation constant of
0.27) when the extra credit is factored out of the final course
grade. In other words, blog readership was not restricted only to
the higher-achieving students; a broad spectrum of students in the
course seemed to participate and find value in the blog.

Next, the average of the value or ``reality link" responses from
the attitudinal survey were analyzed for students who consistently
read and commented on the blog (n=58) and for students who
infrequently or never read or commented on the blog (n=37).
Students from Sections A, B, C, and D comprise both groups.  We
found that although the pre-tests for the two groups were not
statistically different, there was a large difference between the
post-tests of the two groups. The blog-reading group's attitude
did not measurably change over the course of the semester (i.e.
the initial positive attitude was maintained) whereas the non-blog
reading group's attitude deteriorated sharply.  The survey results
and statistics are presented in Table~\ref{tab:5}.

\begin{table}[htp]
\begin{center}
\begin{tabular}{cccc|ccc}
\hline \hline Question & Blog & Non-Blog & Statistical & Blog  &
Non-Blog & Statistical \\
& Pre & Pre & Significance & Post & Post & Significance \\
\hline

1 & 70.8\% &  77.2\% & & 75.5\% & 63.9\% & \\

5 & 61.1\% & 66.6\% & & 66.1\% & 53.9\% & \\

7 & 73.2\%  & 70.6\% & & 75.2\% & 57.9\% &  \\

12 & 84.8\% & 83.2\% & Not & 81.8 \% & 73.4\% & Statistically \\

15 & 70.7\% & 67.6\% & Statistically & 64.7\% & 58.6\% & Significant \\

16 & 69.6\% & 66.3\% & Significant & 61.7\% &  57.9\% & (ES = 1.51)\\

17 & 74.1\% & 71.7\% &  & 68.7\% & 63.2 \% & \\

19 & 66.6\% & 70.8\% & & 65.5\% & 58.9\% & \\

20 & 69.8\% & 64.3\% & & 66.2\% & 57.6\% & \\

24 & 66.7\% & 64.3\% & & 66.4\% & 58.8\% & \\ \hline
Average & 70.4\% & 70.3\% & & 69.2\% & 60.4\% & \\
\hline \hline
\end{tabular}
\end{center}
\caption{Results for all ``reality link" questions for the
attitudinal survey from Semester II (spring 2006); see Appendix I
for the survey questions.  Here the Likert scale survey results
(N=58 for the blogging group and N=37 for the non-blogging group)
were analyzed as interval data with an independent t-test
(comparing pre-tests and post-tests between the blogging and
non-blogging groups).  The results above are averages over the
responses for individual students for each group.  The difference
between the blog and non-blog reading groups was not-statistically
significant for the pre-test but statistically significant with p
$<$ 0.01 for the post-test ($p=.0034$) and an effect size of 1.51.
Scores have been normalized so that 50\% represents a neutral
response on the Likert-style attitudinal survey.} \label{tab:5}
\end{table}

\begin{figure}[htp]
\includegraphics[width=0.69\textwidth]{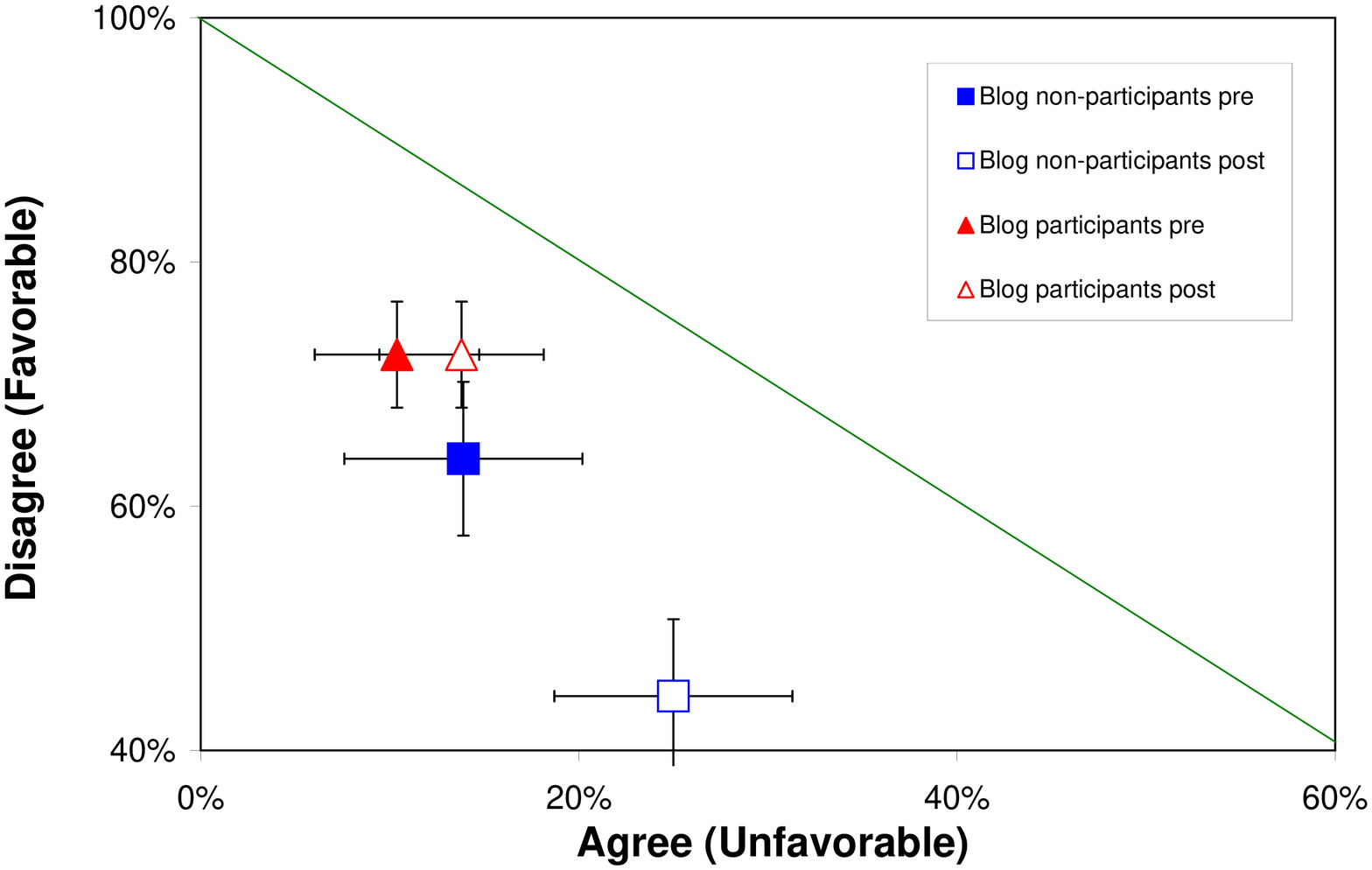}
\caption{A-D plot for Semester II (spring 2006) for question \#7:
``I will/did find it difficult to understand how physics applies
in the real-world" for students who participated in the blog study
and for students who did not read the blog.  Error bars are
1$\sigma$, where $\sigma$ is calculated using Eq.(\ref{eq:2}). In
the binomial analysis a shift of $> 2\sigma$ is considered
statistically significant.} \label{fig:3}
\end{figure}

\begin{figure}[htp]
\includegraphics[width=0.69\textwidth]{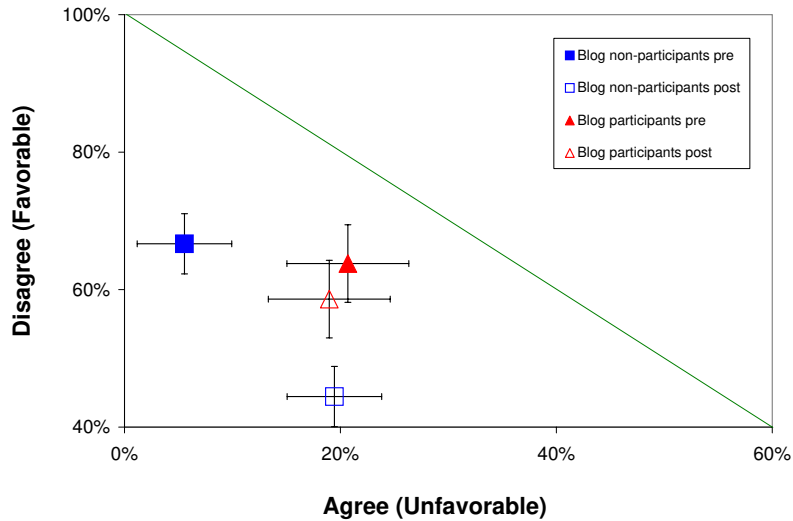}
\caption{A-D plot for Semester II (spring 2006) for question \#19:
``Physics is not useful in my everyday life" for students who
participated in the blog study and for students who did not read
the blog.  Error bars are 1$\sigma$, where $\sigma$ is calculated
using Eq.(\ref{eq:2}). In the binomial analysis a shift of $>
2\sigma$ is considered statistically significant.} \label{fig:4}
\end{figure}

\begin{figure}[htp]
\includegraphics[width=0.69\textwidth]{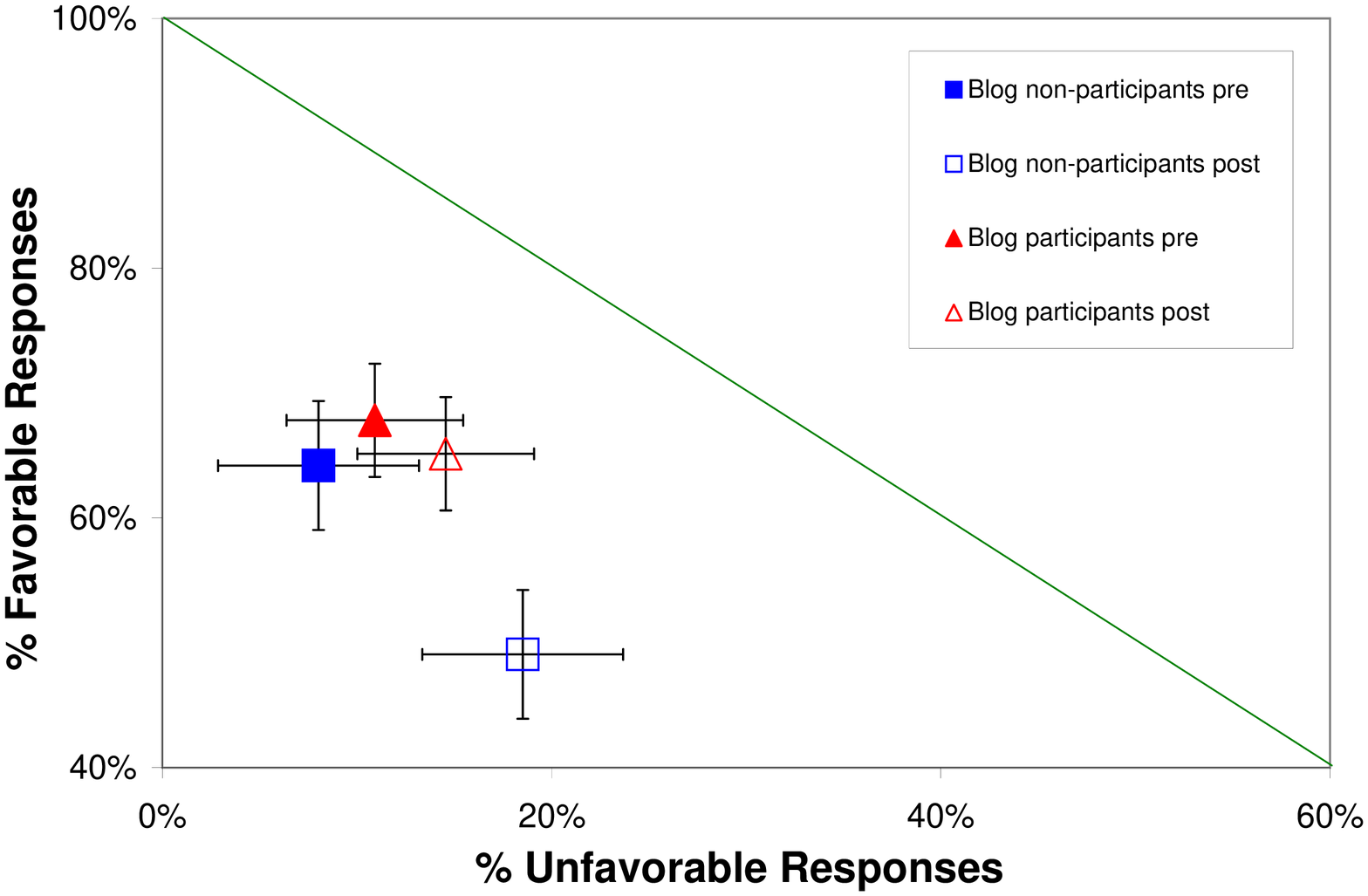}
\includegraphics[width=0.69\textwidth]{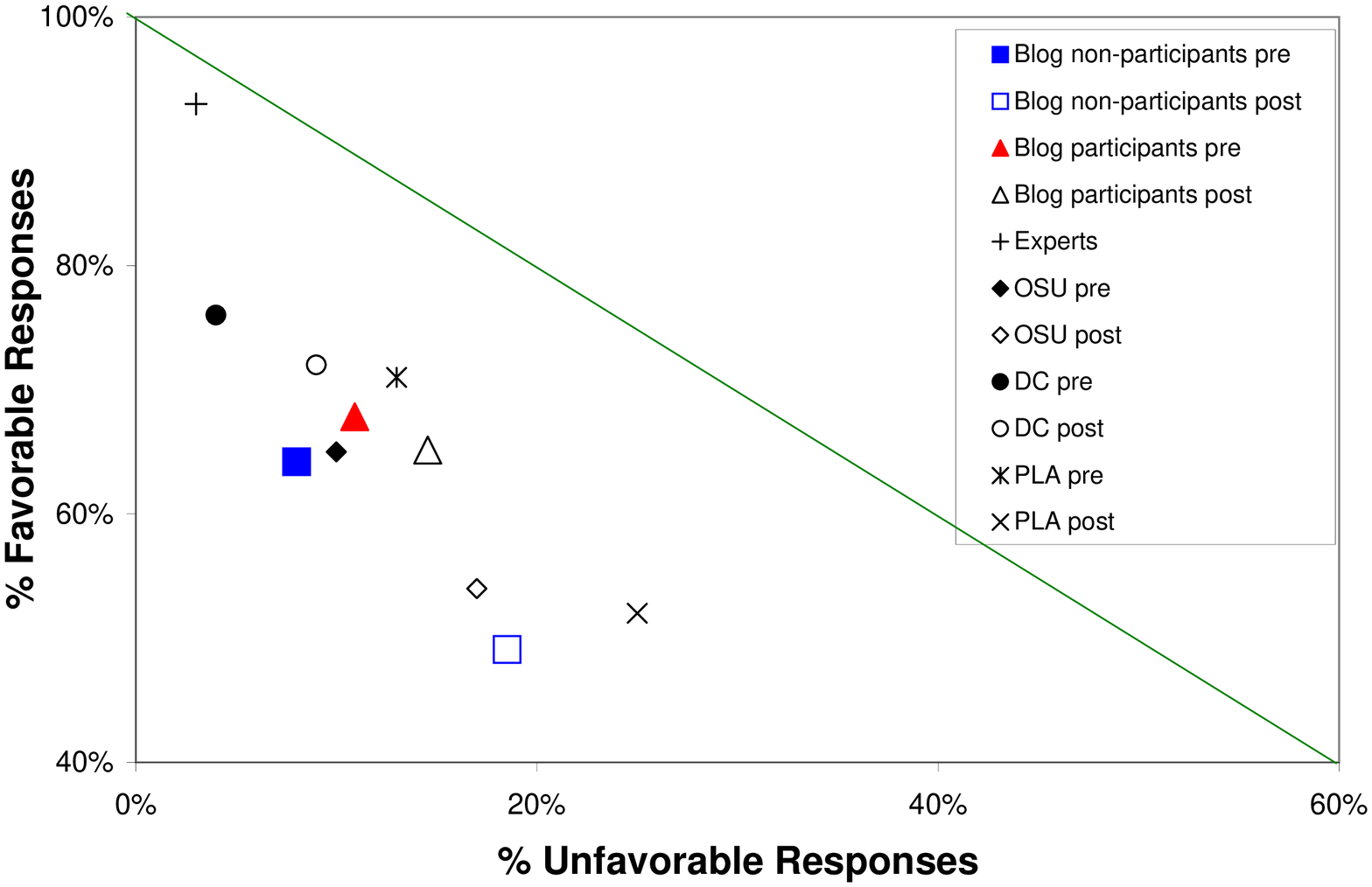}
\caption{Top: A-D plot for Semester II (spring 2006) for an
average of the ``reality link" questions; Bottom: same as above
but with data from the MPEX Survey\cite{MPEX} included for
comparison. Error bars are 1$\sigma$, where $\sigma$ is calculated
using Eq.(\ref{eq:2}). In the binomial analysis a shift of $>
2\sigma$ is considered statistically significant.} \label{fig:5}
\end{figure}

We also analyzed the Semester II (spring 2006) data by utilizing
Agree-Disagree plots.  Rather than presenting plots for all of the
survey questions, we highlight a few examples here. Agree-Disagree
values for both the pre/post attitudinal survey for question \#7
(I will/did find it difficult to understand how physics applies in
the real-world) are plotted as Fig.~\ref{fig:3}. Agree-Disagree
values for Question \#19 (Physics is not useful in my everyday
life) are plotted as Fig.~\ref{fig:4}.  Fig.~\ref{fig:5}a presents
the average of the `reality link' questions.  All of the figures
are plotted for two groups described above (students who read the
blog frequently and those who didn't).  Fig.~\ref{fig:5}b includes
the results from the MPEX ``reality link" questions for Dickinson
College (DC) and a small public liberal arts college (PLA), the
two institutions from the MPEX survey closest in size and
character to Creighton's College of Arts and Sciences.  The expert
response and responses from Ohio State University (OSU), a large
public institution, were included for reference.

In all cases the blog-reading group had little change from pre to
post test.  However, the students who did not read the blog showed
a statistically significant deterioration in their initially
positive attitudes.

\begin{figure}[t]
\includegraphics[width=0.69\textwidth]{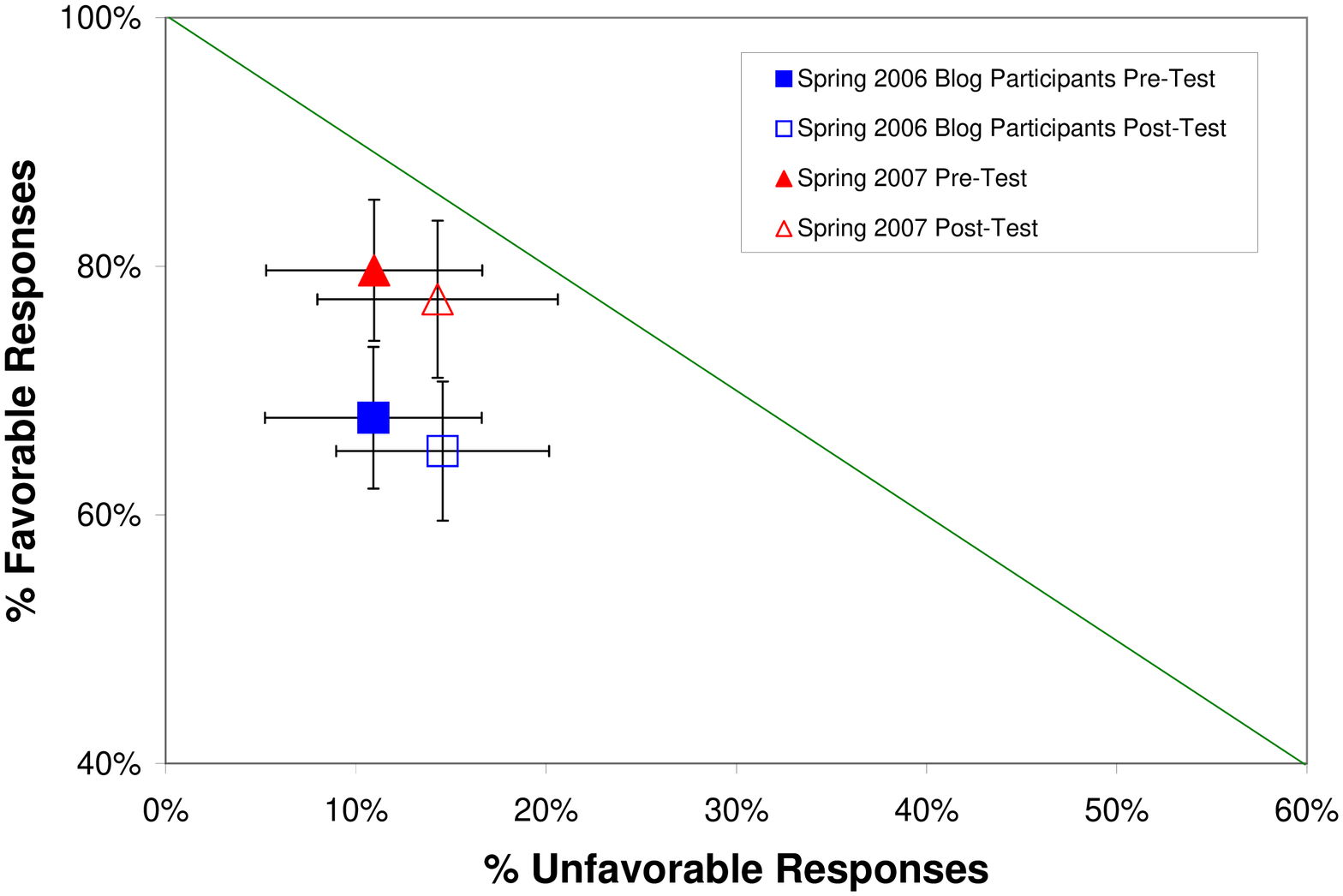}
\caption{A-D plot comparing students who participated in the blog
from Semester III (spring 2006) with students from Semester IV
(spring 2007) in which the blog was a required element of the
course.  Error bars are 1$\sigma$, where $\sigma$ is calculated
using Eq.(\ref{eq:2}). In the binomial analysis a shift of $>
2\sigma$ is considered statistically significant.} \label{fig:7}
\end{figure}

In spring 2007 only one section read the blog and the attitudinal
survey was administered only for this section.  The fact that only
one section using the blog in spring 2007 was not a repudiation of
the blog but rather a chance for other faculty to explore PER-type
work in their own sections.  In this semester the blog was
integrated into the course and was no longer extra-credit; we were
interested to see primarily if students' positive attitudes about
the blog itself would survive if it were no longer extra credit
but rather required.  Because the number of students in the
section was only n=33 and most students fully participated in the
blog, it proved impossible to construct a large enough set to
perform detailed statistics as for Semester II (for example, blog
readership vs. attitude, etc.). However, we have compared the
attitudinal change of the class with previous data.
Fig.~\ref{fig:7} compares the change in the average of the
``reality link" questions from the attitudinal survey for students
who participated in the blog study in Semester II (spring 2006)
and all students in Semester IV (spring 2007) where the blog was
required; in both instances students maintained their initial
positive attitudes, and neither change from pre to post was
statistically significant.  One should be cautious however when
directly comparing Semester II (spring 2006) and Semester IV
(spring 2007) due to the differences in what role the blog played
in the course.

Table~\ref{tab:7} below summarizes the A-D values for Semesters II
and IV (spring 2006 + 2007).  Since questions \#7 and \#19 have
favorable answers which correspond to students disagreeing with
the statement, we have given favorable and unfavorable response
percentages to avoid confusion.

\begin{table}[h]
\begin{center}
\begin{tabular}{llllll}
\hline \hline & Pre-test & & Post-test & & Statistical \\
& Favorable & Unfavorable & Favorable & Unfavorable & Significance \\
\hline
Reality link avergage & & & & & \\
~~~~Spring 2006 - blog & 67.\% & 10.9\% & 65.1\% & 14.5\% & not significant \\
~~~~Spring 2006 - no blog & 64.2\% & 8.0\% & 49.1\% & 18.5\% & significant (shift $\approx 3 \sigma$)\\
~~~~Spring 2007 & 79.7\% & 11.0\% & 77.3\%  & 14.3\% & not significant \\
Q7: I will/did find it difficult to understand & & & & \\
how physics applies in the real-world & & & & \\
~~~~~ Spring 2006 - blog & 72.4\% & 10.3\% & 72.4\% & 13.8\% & not significant\\
~~~~~ Spring 2006 - no blog & 63.9\% & 13.9\%  & 44.4\% & 25.0\% & significant (shift $> 3\sigma$) \\
Q19: Physics is not useful in my everyday life & & & & \\
~~~~~ Spring 2006 - blog & 63.7\% & 20.7\% & 58.6\% & 19.0\% & not significant \\
~~~~~ Spring 2006 - no blog & 66.7\% & 5.6\%  & 44.4\%  & 19.4\% & (shift $> 5 \sigma$)\\
\hline \hline
\end{tabular}
\end{center}
\caption{Student A-D values for plotted spring 2006 + 2007 items.
The Likert scale survey results were analyzed as ordinal data here
using the binomial agree/disagree analysis presented in the MPEX
paper. $\sigma$ is calculated using Eq.(\ref{eq:2}); a shift of $>
2\sigma$ is considered statistically significant.} \label{tab:7}
\end{table}

\section{Student Reactions}

Student reaction to the blog, even in Semester IV when it was no
longer extra-credit, was overwhelmingly positive. For example, for
spring 2006 and spring 2007 (Semesters II and IV) the vast
majority of students found the blog interesting and were more
interested in physics because of the blog.  Student reactions to
the blog are summarized below in Table~\ref{tab:6}.  What is
particularly striking is the fact that there was no large scale
deterioration in the generally positive feeling for the blog in
Semester IV (spring 2007) as compared to spring 2006 (Semester
II); in Semester IV the blog was mandatory and students' posts
counted as credit towards their final grade.  In other words, the
blog was popular with students even when it ceased to be extra
credit.

\begin{table}[h]
\begin{center}
\begin{tabular}{lllll}
\hline \hline & Spring & 2006 & Spring & 2007 \\
Attitudinal Survey Question & Agree & Disagree & Agree & Disagree \\
\hline The blog made the class more interesting to me & 80.2\% & 19.8\% & 73.3\% & 26.7\%\\
The blog helped improve my learning in this class & 60.3\% & 39.7\% & 52.2\% & 47.8\%\\
I am more interested in physics due to the blog & 68.7\% & 31.3\% & 66.7\% & 33.3\% \\
I am glad we had the course blog & 85.4\% & 14.6\% & 65.2\% & 34.8\% \\
The course blog would have been valuable & & & & \\
even if we didn't earn (extra) credit & 50.7\% & 49.3\% & 65.2\% & 34.7\% \\
\hline \hline
\end{tabular}
\end{center}
\caption{Student Post-test reactions to the course blog in Spring
2006.  The 5-pt Likert data were fit to a binomial agree-disagree
scale as described earlier.} \label{tab:6}
\end{table}

In each semester a written blog questionnaire was distributed at
the end of the semester to give students a chance to give written
feedback on their blog experience.  Students generally had very
positive feedback about the blog; we give some example student
responses below.

\begin{quote}
``It made it easier to appreciate physics.  It's so boring talking
about blocks and strings and bike wheels, but talking about
planets and bigger things was great." -- Student \#1 \end{quote}

\begin{quote}
``It made me aware of just how much physics impacted our everyday
lives.  It also made me realize that physics is in fact useful in
my profession (and many more) when I originally couldn't see the
connection." -- Student \#2 \end{quote}

\begin{quote}
``I'm a little bit more interested [in physics], and the blog
definitely helped!  It was fun and interesting and true to life,
not the `ideal situation' physics class deals with." -- Student
\#3 \end{quote}

\begin{quote} ``It is often hard to see physics at work in
everyday life, especially compared to biology.  The blog gave me a
greater appreciation for physics and motivated me to do my
homework!" -- Student \#4 \end{quote}

\begin{quote} ``The blog made me feel like I could interact with
the class more.  At 8:30 am you don't feel like getting involved!"
-- Student \#5 \end{quote}

Despite the fact that the blog entailed time outside of class,
students enjoyed reading the posts, commenting and asking
questions, and being able to interact with their classmates in a
new way.  Many students were passionate supporters of the blog,
going as far to say that it was their favorite component of the
course.  A common theme which emerged was that students
appreciated the chance to see how physics was applied to the real
world, and how physics had a great deal to say about systems far
more complex than those dealt with in class.

\section{Conclusions}

In this paper, we have explored the effect of students'
participation in a course weblog on their overall attitude towards
the value of physics principles learned in the introductory
physics classroom.  The course blog featured postings about
various modern applications of physics in areas of technology and
nature, and encouraged students to conduct further research and
exploration, discuss with other students, and pose questions. This
was designed to provide the students with a real world vision of
physics, rather than one that revolves around the more mundane and
necessarily simple examples that can be covered in class.

Students' attitudes and perceptions of the importance and value of
physics were examined using a 5-pt Likert scale attitudinal survey
which was given as a pre and post test for four semesters of
introductory physics. From the data gathered using this
attitudinal survey, we can answer the main question raised in this
paper: can the use of a semester-long blog have a significant,
positive impact on how students feel about the relevance of
physics? Looking at the average responses of the ``reality link"
questions on the survey, we see that those students who did not
read the blog in semesters I and III had a statistically
significant deterioration in attitude over the course of the
semester, while those who did participate in the blog in each of
the four semesters showed little, if any, change; in other words,
the initial optimistic attitudes students tended to start the
course with remained.  When compared with the results of other
universities on the MPEX ``reality link" items, we can see that
the students at CU who did not participate in the blog are
comparable to those from other institutions.  In a more explicit
manner, students in semesters II and IV were asked if they
personally felt that the blog improved their overall experience in
general physics.  On each of these questions, over half (ranging
from 50.7\% to 85.4\%) of the students responded in a positive
manner. The students felt that overall, the blog was helpful in
learning the material covered in the classroom, made physics more
interesting, and was generally enjoyable.  These encouraging
results reflect the sustained positive attitude of blog
participants on the more indirect ``reality link" questions.

In the introduction we detailed four potential benefits of
blogging from Ferdig and Trammel \cite{blogs6}; while writing and
maintaining the blog we saw these benefits become tangible to our
students.  In writing comments to posts on the blog students were
forced to do outside research, learning about areas of physics
that weren't covered in class; for example, in the Little Green
Men post reproduced as Appendix II, students researched a bit
about the life and death of stars, stellar nucleosynthesis, and
black holes.  Table \ref{tab:6} shows that the blog made class
more interesting for the vast majority of students, and that
students were more interested in physics due to the blog.  Time
and time again we saw students who never participated or asked
questions in class being very vocal and interactive on the blog;
students felt comfortable asking questions and often (instead of
waiting for the instructor) answered other students' questions.
And finally the blog got students thinking about and discussing
physics outside the classroom.  SiteMeter \cite{sitemeter}
statistics show that most blog comments were posted late into the
night when students are most active and which is outside of the
traditional hours of instruction and instructor-student
interaction.

Finally, because attitude and learning go hand-in-hand, and
because of the continuously increasing use of technology (and
therefore physics) in our world today, it is essential to explore
why perceptions of the value of physics generally become more
negative after a semester of instruction and how to combat this.
In addition, studies have shown that undergraduates are
withdrawing from majors in the physical sciences despite
demonstrating an ability to succeed in such courses\cite{Seymour};
this fact should be alarming to any instructor in the sciences
making student attitudes even more important.  If students cannot
make solid connections between physics learned in the classroom
and how the world works, they will probably lose interest and
leave the semester questioning, ``Why was this important?"  Or, in
extreme cases, this negative attitude may drive undergraduate
physics majors to choose other areas of study.

With these many concerns in mind, we hope to have supplied an
example of how to apply a blog to an introductory physics class as
an effective way to study and improve at least a subset of student
attitudes (the real-world connections of physics), as well as to
help lay the foundation for further research into this topic.

\section{Acknowledgements}

G.D. would like to thank his colleague Janet Seger for running the
course blog and collecting student attitudinal data during Fall
2006 when he was not teaching general physics due to a scheduled
junior-faculty course release.  G.D. would also like to thank
Jennifer Johnson, a student in general physics, for useful
discussions regarding the blog.

\newpage

\section{Appendix I: The Attitudinal Survey}

\subsection{Survey Items}

In the table below we give the complete list of items for the
attitudinal survey.  Starred items, we believe, correspond well
with the MPEX ``reality link" category or the ``real world
connections" category on the CLASS inventory.  The items denoted
with a $\dagger$ are the items included in the averages over
``reality link"-type questions; these were the items we felt most
directly related to the blog and which had the highest internal
consistency and reliability.

\begin{table}[h]
\begin{center}
\begin{tabular}{|l|l|}
\hline & \textbf{Survey Item}  \\ \hline
\hline 1$\dagger$ & Physics is irrelevant to my life$^\star$\\
\hline 2 & I will (did) understand how to apply analytical
reasoning to
physics \\
\hline 3 & I will have (had) no idea what's going this semester in
physics \\
\hline 4 & I will (did) like physics this semester \\
\hline 5$\dagger$ & What I learn in physics this semester will not
be
useful in my career$^\star$ \\
\hline 6 & Physics is highly technical \\
\hline 7$\dagger$ & I will (did) find it difficult to understand
how
physics applies in the real-world$^\star$ \\
\hline 8 & I will (did) enjoy taking this physics course \\
\hline 9 & Physics is purely memorizing a massive collection of
facts and formulas \\
\hline 10 & Physics is a complicated subject \\
\hline 11 & I can (did) learn physics \\
\hline 12$\dagger$ & Physics is worthless \\
\hline 13 & I see and understand physics in technology and the
world around me$^\star$ \\
\hline 14 & I am scared of physics \\
\hline 15$\dagger$ & Skills I learn in physics will make me more
employable$^\star$ \\
\hline 16$\dagger$ & I can use physics in my everyday life$^\star$ \\
\hline 17$\dagger$ & Physics is not applicable to my life outside
school$^\star$ \\
\hline 18 & Physics should be a required part of my professional
training \\
\hline 19$\dagger$ & Physics is not useful in my everyday life$^\star$ \\
\hline 20$\dagger$ & I will (do) look at the world differently
after taking this class$^\star$ \\
\hline 21 & Physics is useful to science or medical professionals
in all fields$^\star$ \\
\hline 22 & What I learn will be applicable to my life outside my
job$^\star$ \\
\hline 23 & Physics should be required as part of my professional
training$^\star$ \\
\hline 24$\dagger$ & Physics this semester will (did) make me
change some of my ideas about how the world works$^\star$ \\
\hline 25 & This semester of physics will be (was) interesting \\
\hline 26 & I am glad I am taking (took) this class \\
\hline
\end{tabular}
\end{center}
\label{tab:8}
\end{table}

\subsection{Reliability}

The attitudinal survey we use in this study grew out of a
statistics instrument (measuring the attitude of statistics
students towards that topic) and was later employed by Zeilik et
al. \cite{Zeilik} in astronomy courses at the University of New
Mexico.  Since this survey has not been as rigorously utilized and
tested as the MPEX and CLASS surveys we performed several tests to
demonstrate the reliability and validity of the attitudinal
survey.

To test the reliability of the survey we calculated a
Cronbach-alpha value (a measure for the internal consistency of
the instrument) for the reality link questions on the attitudinal
survey.  Table \ref{tab:9} gives the Cronbach-alpha values for the
reality link items in the instrument for various semesters.

\begin{table}[h]
\begin{center}
\begin{tabular}{lll}
\hline \hline Semester & Number & $\alpha$ value \\
\hline Spring 2006 & (n=94) & 0.85 \\
Spring 2007 & (n=28) & 0.70 \\
Spring 2008 & (n=78) & 0.87 \\
\hline Average & (n=200) & 0.84 \\
\hline \hline
\end{tabular}
\end{center}
\caption{Cronbach-alpha values for the reality link cluster
questions in the attitudinal survey.  Spring 2008 data (not used
elsewhere) is included only to gain further confidence in the
reliability of the instrument.} \label{tab:9}
\end{table}

In comparison, the MPEX survey has an alpha of 0.67 for the
reality link cluster; the overall MPEX survey has a Cronbach Alpha
of .806.  In this Ph.D. thesis detailing the development of the
MPEX instrument, Saul uses the value of 0.7 as the lower limit for
an Alpha for a test that can be considered reliable \cite{Saul}.

In spring 2008 an essentially random group of students (n=26) in
general physics were given both the attitudinal survey and the
CLASS instrument (in essence to test the validity of the
attitudinal survey).  Although this group is not included in this
study (due to different methodologies and a different phase of the
blog study), we wanted to determine if the attitudinal survey we
use gives similar results to a well-tested survey like the CLASS.
Table (\ref{tab:10}) gives the correlation coefficients between
the ClASS and attitudinal surveys.

\begin{table}[h]
\begin{center}
\begin{tabular}{lc}
\hline \hline Cluster & Correlation Coefficients \\
\hline Reality Link & 0.70 \\
Personal Interest & 0.72 \\
Problem Solving Confidence & 0.25 \\
 \hline \hline
\end{tabular}
\end{center}
\caption{Correlation coefficients between the results of the
attitudinal survey and the CLASS instrument for questions
corresponding to three separate clusters (n=28, spring 2008).}
\label{tab:10}
\end{table}

Although this sample is of limited size, we can conclude that
students' responses to reality link questions correlate strongly
between the attitudinal survey and the CLASS instrument.  This
gives increased confidence that both instruments measure students'
attitudes towards the ``connection between physics and reality -
whether physics is unrelated to experiences outside the classroom
or whether it is useful to think about them together" as defined
by the MPEX survey \cite{MPEX}.

\newpage

\section{Appendix II: Sample Blog Post}

Below we include a sample blog post used in Semesters I and III
relating to angular momentum as well as several student comments.
Pictures and hyperlinks have been removed.

\subsection{Blog Post: Little Green Men and Pulsars}

For today's blog entry I want to share an interesting example of
conservation of angular momentum. Our story starts in 1967, where
graduate student Jocelyn Bell was constructing a large radio
telescope (at 81.5 MHz) at Cambridge University in England.  The
radio telescope consisted of receivers mounted on tall poles
(about 9 feet tall), strung together by wires, with the whole
assembly covering several acres of land. Bell's job was to monitor
the output of the radio telescope; this meant painstakingly going
through about 250 m of charts that were produced every four days.
Bell noticed a strange signal in the data which she at first
attributed to noise or interference. However, upon closer
examination it appeared that the signal was not noise at all, but
a repeating signal coming from outside the solar system (the
signal source moved at the same rate as stars in the sky, showing
it was not a manmade signal nor a signal coming from nearby).  The
signal repeated itself every 1.339 seconds and was very regular.
Bell and her advisor, Anthony Hewish, named the source of the
signal LGM-1, or Little Green Men-1; it was seriously thought at
the time that this might be the first detection of an
extraterrestrial signal.  Hewish and Bell published their results
in the journal Nature, which sparked lively debate in astronomical
circles about the source of the signals.  Eventually other
pulsating radio sources like the one discovered by Bell were
found, which convinced astronomers that they were looking at a
natural, not extra-terrestrial signal.

So, what created the signal Jocelyn Bell discovered?  To
understand what she had found we need to learn a little bit about
stars and stellar evolution.  It turns out that stars are not
eternal, but are born and die just like everything else in the
Universe.   Stars spend most of their lives (about 20 billion
years for a star like our Sun) burning hydrogen in their core into
helium through the process of nuclear fusion (putting four
hydrogen nuclei together to form a helium nucleus releases
tremendous energy).   The energy released by fusion is transported
through the star and eventually reaches us on Earth.  During its
whole lifetime, a star is in a state of careful balance.  Energy
generation through nuclear fusion tends to expand the star while
gravity tends to try to collapse the star.  As long as the star
can generate energy through fusion, gravity and collapse are held
at bay.

However, very massive stars (stars like Betelgeuse with masses 10x
or so of our Sun's mass) can eventually run out of hydrogen in the
core to burn and produce energy.  A series of run-away reactions
occurs, and the stars begins to burn heavier and heavier elements
in its interior. Finally after burning hydrogen to helium, then
helium to carbon, and so on, iron is formed in the core.  Iron is
a special element in that in order to fuse iron you need to add
energy rather than getting energy out (it's an endothermic
reaction rather than an exothermic).  So the star can no longer
forestall gravitational collapse by producing energy in the core.
The star collapses and then explodes in what astronomers call a
supernova.  Even though a galaxy like our Milky Way contains about
100 billion stars, one supernova can briefly outshine the entire
galaxy.

Although you might think that the star is completely obliterated
in such an explosion, the core of the star actually remains.  The
core has been so squeezed by gravity that electrons and protons
have combined to form neutrons.  The leftover core is called a
neutron star.  Our sun's diameter is about $1.3 \times 10^6$ km,
but a neutron star is only about 25 km in size.  The gravity of a
neutron star is so strong (because of its mass and its small size)
that the acceleration of gravity on the surface is not 9.8 m/s$^2$
but about $10^{11}$ m/s$^2$.

So what does this have to do with the pulsating radio objects (or
pulsars) that Jocelyn Bell discovered?  Well, Bell's pulsars are
actually neutron stars that are rotating very, very rapidly.  In
fact, the rotational speeds of neutron stars are almost
unbelievable.  So why do they rotate so quickly?  Actually it
comes down to simple conservation of angular momentum; as a star
collapses into a supernova, the angular momentum of the star must
remain conserved.  For example, a star like our Sun has a angular
speed of about $2 \times 10^{-6}$ rad/sec (it rotates once in
25.38 days) and a radius of about $10^6$ km.  If our sun were to
collapse to a neuron star with a radius of only 25 km, we could
use conservation of angular momentum to find the final rotational
velocity.  We know that

\[ I_i \omega_i = I_f \omega_f \]

\noindent And since the moment of inertia for a sphere is 2/5
MR$^2$, we can solve for the final angular velocity:

\[ \omega_f = (R^2_i/R^2_f) \omega_i \]

\noindent Plugging in numbers gives $\omega_f \approx 8700$
rad/sec or a rotational period of about $7 \times 10^{-5}$ sec.
That's pretty darn fast! And since rotational kinetic energy goes
like $\omega^2$, a rotating neutron star, for such a small object,
packs quite a lot of rotational energy.

The reason that pulsars or rotating neutron stars give off energy
as radio pulses is that some of the rotational kinetic energy is
being converted to radio energy as the pulsar slows down.  As a
pulsar or neutron star rotates, thin beams of radiation speed
through space, like the beam of a lighthouse.  It was these beams
that Jocelyn Bell and the radio telescope at Cambridge discovered.
For more information about pulsars I suggest starting at
Wikipedia's entry here.

So, to conclude, what I find fascinating about physics is that a
simple experiment like sitting on a rotating stool and watching
your rotational speed change as you bring your arms in and out can
actually explain the workings of something as incredible as a
rotating neutron star.  Physics done in the lab actually controls
how things work out in the Universe.  So when we say conservation
of angular momentum is a fundamental principle we really, really
mean it!

\subsection{Sample Student Comments}

\begin{quote}
``This was so interesting! We always learn in class that the laws
of physics do not change...but often, it does not occur to me that
they operate in something as small as a molecule or something as
big as a star! Those little neutron stars rotate so fast and have
so much energy in them that even though they are so far away we
can still see the effects here on Earth, as Jocelyn Bell
discovered. The black hole thing is interesting too...I am
wondering- how did scientists figure out the mass at which the
gravity would cause the star to collapse? This phenomenon does not
occur anywhere else in nature, so how can we possibly know how it
happens? Amazing!" -- Student \#1 \end{quote}

\begin{quote}
``What I found most interesting doesn't really have anything to do
with angular momentum.  I am curious to how one supernova can
briefly outshine the entire galaxy, even though a galaxy like the
Milky Way contains about 100 billion stars. That is a lot of
outshining to do! So I did a little research and didn't find an
answer but I did find this - supernova explosions are the main
source of all the elements heavier than oxygen, and they are the
only source of many important elements. For example, all the
calcium in our bones and all the iron in our hemoglobin were
synthesized in a supernova explosion, billions of years ago.  Who
would have thought?" -- Student \#2 \end{quote}

\begin{quote}
``I agree with [another student]; this is by far the most
interesting blog post yet! After my instructor mentioned neutron
stars in class (and did some of the angular momentum equations) I
became very fascinated and did a little research on my own. I
learned that neutron stars are about the size of Manhattan island,
but are more massive than the sun. I also learned that if you
could somehow get your hands on a tablespoon-sized piece of a
neutron star it would weigh a billion tons. Pretty tough to get
your head around huh? Some of the sites I visited noted something
about these neutron stars being superfluid; I was under the
impression that they were nearly solid iron....any clarification?"
-- Student \#3
\end{quote}


\begin{thebibliography}{99}

\bibitem{McDermott} L.C. McDermott, ``Millikan Lecture 1990: What
we teach and what is learned - closing the gap," Am. J. Phys.
\textbf{59}, 301-315 (1991).

\bibitem{Redish1} E. Redish, ``Millikan Lecture 1998: Building a
Science of Teaching Physics," Am. J. Phys. \textbf{67}, 562-573
(1999).

\bibitem{Redish2} E. Redish and R. Steinberg, ``Teaching Physics:
Figuring out what works," Phys. Today \textbf{52} (1), 24-30
(1999).

\bibitem{Redish3} L.C. McDermott and E. Redish, ``Resource Letter: PER-1: Physics Education
Research,'' Am. J. Phys. \textbf{67}, 755-767 (1999).  Although a
few years old, this resource letter contains a wonderful
assortment of PER references and reviews.

\bibitem{Mazur} C. Crouch and E. Mazur, ``Peer Instruction: Ten
years of Experience and Results," Am. J. Phys. \textbf{69},
970-977 (2001).

\bibitem{JITT} G. Novak, E. Patterson, A. Gavrin, and W.
Christian, \textit{Just-in-Time Teaching} (Prentice-Hall, Upper
Saddle River, NJ 1999).

\bibitem{UofW} L.C. McDermott, P.S. Shaffer \textit{et al}.,
\textit{Tutorials in Introductory Physics} (Prentice-Hall, Upper
Saddle River, NJ 1998).

\bibitem{workshop} P. Laws \textit{et al}., \textit{Workshop Physics Activity Guide (Second Edition)},
Modules 1-4 with Appendices (John Wiley and Sons, New York, 2004).

\bibitem{MPEX} E. Redish, J. Saul, and R. Steinberg, ``Student
expectations in introductory physics," Am. J. Phys. \textbf{66},
212-224 (1998).

\bibitem{Zeilik} M. Zeilik, C. Schau, and N. Mattern, ``Conceptual
Astronomy. II. Replicating conceptual gains, probing attitudde
changes across three semesters," Am. J. Phys. \textbf{67}, 923-927
(1999).

\bibitem{Zeilik2} M. Zeilik \textit{et al}., ``Conceptual astronomy: A novel model
for teaching postsecondary science courses," Am. J. Phys.
\textbf{65}, 987-996 (1997).

\bibitem{CLASS} W. Adams, K. Perkins, N. Podolefsky, M. Dubson, N. Finkelstein, and C. Wieman, ``New instrument
for measuring student beliefs about physics and learning physics:
The Colorado Learning Attitudes about Science Survey," Phys. Rev.
ST Phys. Educ. Res. \textbf{2}, 010101 (2006).

\bibitem{stats1} I. Gal and L. Ginsburg, ``The Role of Belief and
Attitudes in Learning Statistics: Towards and Assessment
Framework," J. Stat. Ed. \textbf{2} (2), p. (1994)

\bibitem{stats2} C. Schau, J. Stevens, T. Dauphinee, and A. Del
Vecchio, ``The Development and Validation of the Survey of
Attitudes Towards Statistics," Educ. and Pysch. Measurement
\textbf{55}, 868-875 (1995).

\bibitem{chem1} J. Dalgety, R. Coll, and A. Jones, ``Development of
Chemistry Attitudes and Experiences Questionnaire (CAEQ)," J. Res.
Sci. Teach. \textbf{40} (7), 649-668 (2003).

\bibitem{chem2} C. Anders and R. Berg, ``Factors related to observed attitude
change toward learning chemistry among university students," Chem.
Ed. Res. and Prac. \textbf{6} (1), 1-18 (2005).

\bibitem{Schoenfeld} A. Schoenfeld, ``Learning to think
mathematically: Problem solving, metacognition, and sense-making
in mathematics," in \textit{Handbook of Research in Mathematics
Teaching and Learning}, edited by D.A. Grouws (MacMillan, New
York, 1992), 334-370.

\bibitem{Koballa} T. Koballa and F. Crawley, ``Attitude-Behavior
Change in Science Education: Part II, Results of an Ongoing
Research Agenda," Paper presented at the Annual Meeting of the
National Association for Research in Science Teaching (65th,
Boston, MA, March 21-25, 1992).

\bibitem{FCI} I. Halloun and D. Hestenes, ``The Initial Knowledge State of College Physics Students",
Am. J. Phys. 53, 1043-1055 (1985 and ``Common Sense Concepts about
Motion", Am. J. Phys. \textbf{53}, 1056-1065 (1985).

\bibitem{Coletta} V.~Coletta and J.~Phillips, in ``FCI Normalized Gains and Population
Effects" a talk given at the Summer 2005 AAPT Conference available
at http://myweb.lmu.edu/jphillips/PER/AAPT-summer05.pdf.

\bibitem{blogs1} R. Blood, ``Weblogs: A history and perspective,"
http://www.rebeccablood.net/essays/weblog\_history.html.  We find
it only fitting to point the reader to an actual blog for a
discussion of blogs in general.

\bibitem{blogs2} S. Krause, ``Blogs as a Tool for Teaching," The
Chronicle of Higher Education \textrm{51} Issue 42, B33 (2005).

\bibitem{blogs3} E. Brownstein \textit{et al}., ``An educator's guide to blogs," Conference presentation at
National Educational Computing Conference, Philadelphia, June 27-3
(2005).

\bibitem{blogs4} T. Embrey, ``You Blog, We Blog," Teacher
Librarian \textbf{30} (2), 7-9 (2002).

\bibitem{blogs5} http://blogtalk.net.  Blog Talk is a yearly
international conference held in Vienna, Austria, which focuses on
the role of blogs in education.   The website includes videos of
the talks at the most recent 2006 conference.

\bibitem{blogs6} R. Ferdig and K. Trammell, ``Content Delivery in the 'Blogosphere',"
T.H.E. Journal; http://www.thejournal.com/articles/16626

\bibitem{blogs7} ``Blogs move student learning beyond the
classroom: An interview with Alex Halavais," Online Classroom
Dec., 4-8 (2004).

\bibitem{blogs8} E. Brownstein and R. Klein, ``Blogs: Applications
in Science Education,'' J. Col. Sci. Teach. \textbf{35} v. 6,
18-22 (2006).

\bibitem{Good For} A. Gavrin and G. Novak, ```What is Physics Good For?'  Motiving
students with online materials," Proceedings of the IASTED
Internal Conference Computers and Advanced Technology in
Education, May 6-8, 1999, Philadelphia, Pennsylvania, USA; the
article can be found online at
http://webphysics.iupui.edu/JITT/CATE1999.doc

\bibitem{FLAG} http://www.flaguide.org; FLAG is the Field-tested
Learning Assessment Guide for science, math, engineering, and
technology instructors.  The FLAG website has a compilation of
Classtroom Assessment Techniques (CATs) which can be matched up
with the instructors desired goal.

\bibitem{Kortemeyer} G. Kortemeyer, ``Correlations between student
discussion behaviour, attitudes, and learning," Phys. Rev. ST
Phys. Educ. Res. \textbf{3}, 010101-010109 (2007).

\bibitem{Schau} C. Schau, J. Stevens, T.L. Dauphinee, and A. Del
Vecchio, ``The development and validation of the Survey of
Attitudes Towards Statistics," Ed. Psych. Measurement \textbf{55},
868-875 (1995).

\bibitem{cugenphys} http://www.physics212.blogspot.com/

\bibitem{YouTube} http://www.youtube.com

\bibitem{blogger} http://www.blogspot.com

\bibitem{haloscan} http://www.haloscan.com

\bibitem{wordpress} http://www.wordpress.com

\bibitem{howstuffworks} http://www.howstuffworks.com

\bibitem{VASS} I. Halloun and D. Hestenes, ``Interpreting VASS
Dimensions and Profiles for Physics Students," Science and
Education \textbf{7}, 553-557 (1998).

\bibitem{Saul} J. Saul, ``Beyond problem solving: Evaluating introductory
physics courses through the hidden curriculum," Ph.D. Thesis,
University of Maryland (1998); available online at
http://www.physics.umd.edu/perg/dissertations/Saul/.

\bibitem{Seymour} E. Seymour, ``Guest comment: Why undergraduates leave
the sciences," Am. J. Phys. \textbf{63}, 199-202 (1995).

\bibitem{DIRECT} P. Engelhardt, and R. Beichner, ``Students' understanding
of direct current resistive electrical circuits", Am. J. Phys.
\textbf{72}, 98-115 (2004).

\bibitem{CSEM} D. Maloney, T. O'Kuma, C. Hieggelke, and A. Van Heuvelen,
``Surveying students' conceptual knowledge of electricity and
magnetism", Am. J. Phys. \textbf{69}, S12 (2001).

\bibitem{sitemeter} http://www.SiteMeter.com


\end{thebibliography}
\end{document}